\documentclass[10pt]{iopart}

\usepackage{url}
\usepackage{bm,bbm}
\usepackage{graphicx}
\usepackage{color}
\usepackage[colorlinks=true, urlcolor=blue, linkcolor=blue, citecolor=blue, pdftex]{hyperref}
\usepackage[english]{babel}
\usepackage{soul}
\usepackage{blindtext}

\newcommand{\dagga}{{\phantom{\dagger}}}

\usepackage{iopams}  
\begin{document}

\title{Dynamical properties of N\'eel and valence-bond phases in the $J_1-J_2$ model on the honeycomb lattice}

\author{Francesco Ferrari$^{1,2}$ and Federico Becca$^3$}

\address{
$^{1}$ Institute for Theoretical Physics, Goethe University Frankfurt, Max-von-Laue-Stra{\ss}e 1, D-60438 Frankfurt a.M., Germany

$^{2}$ SISSA-International School for Advanced Studies, Via Bonomea 265, I-34136 Trieste, Italy

$^{3}$ Dipartimento di Fisica, Universit\`a di Trieste, Strada Costiera 11, I-34151 Trieste, Italy
}

% \ead{submissions@iop.org}
\vspace{10pt}

\begin{abstract}
By using a variational Monte Carlo technique based upon Gutzwiller-projected fermionic states, we investigate the dynamical 
structure factor of the antiferromagnetic $S=1/2$ Heisenberg model on the honeycomb lattice, in presence of first-neighbor ($J_1$) 
and second-neighbor ($J_2$) couplings, for ${J_2 < 0.5 J_1}$. The ground state of the system shows long-range antiferromagnetic 
order for ${J_2/J_1 \lesssim 0.23}$, plaquette valence-bond order for ${0.23 \lesssim J_2/J_1 \lesssim 0.36}$, and columnar dimer 
order for ${J_2/J_1 \gtrsim 0.36}$. Within the antiferromagnetic state, a well-defined magnon mode is observed, whose dispersion 
is in relatively good agreement with linear spin-wave approximation for $J_2=0$. When a nonzero second-neighbor super-exchange is 
included, a roton-like mode develops around the $K$ point (i.e., the corner of the Brillouin zone). This mode softens when $J_2/J_1$ 
is increased and becomes gapless at the transition point, $J_2/J_1 \approx 0.23$. Here, a broad continuum of states is clearly 
visible in the dynamical spectrum, suggesting that nearly-deconfined spinon excitations could exist, at least at relatively high 
energies. For larger values of $J_2/J_1$, valence-bond order is detected and the spectrum of the system becomes clearly gapped, 
with a triplon mode at low energies. This is particularly evident for the spectrum of the dimer valence-bond phase, in which the 
triplon mode is rather well separated from the continuum of excitations that appears at higher energies.
\end{abstract}

\ioptwocol

\section{Introduction}

The Heisenberg model represents the simplest playground to investigate thermal and quantum phase transitions, possibly leading to phases 
with unconventional properties, including topological order and fractional excitations. Thermal fluctuations have been considered in 
presence of a large degeneracy within the low-energy manifold, in connection to the order-by-disorder mechanism;~\cite{Moessner2006} 
by contrast, the effect of quantum fluctuations at zero-temperature has been widely considered to drive a magnetically ordered ground 
state into a disordered phase.~\cite{Imai2016} By varying the geometry of the underlying lattice, the value of the spin $S$, and the 
range of the super-exchange interactions, an overabundance of potentially interesting scenarios has been proposed and investigated in 
the previous 30 years.~\cite{Zhou2017} In addition, recent studies pointed out the relevance of perturbing terms that breaks the spin 
rotational symmetry, to stabilize quantum states with intriguing physical properties.~\cite{Maksimov2019}

As far as the quantum systems are concerned, there has been a huge effort to detect and characterize the so-called spin liquids,
which represent a class of states with no {\it local} broken symmetry whatsoever and elementary excitations that carry fractional
quantum numbers of the original constituents (i.e., in a $S=1/2$ model, a spin-flip excitation carries $S=1$ and, therefore, an 
excitation with $S=1/2$ has a fractional quantum number).~\cite{Savary2016} Quantum spin liquids emerge from magnetic frustration, 
due to the existence of competing super-exchange couplings, and are characterized by a strong entanglement between spins on distant
sites. A paradigmatic representation can be obtained by the so-called resonating-valence bond picture, introduced by Anderson and 
Fazekas~\cite{Anderson1973,Fazekas1974}: here, each spin is coupled to a partner to form a singlet, thus defining a particular 
``singlet covering'' of the entire lattice; then, an exponentially large linear combination of such coverings is taken to obtain 
a fully-symmetric quantum state. This approach generalizes the idea of valence-bond resonance~\cite{Pauling1933} introduced for 
the Benzene molecule to the case of singlet coverings of a lattice. Two broad classes of spin liquids exist: gapped and gapless. 
In the former one, the probability of having a singlet at large distance decays exponentially, while in the latter one the decay 
is algebraic (thus leading to a much more entangled state). Fully gapped spin liquids represent stable phases of matter; instead,
gapless spin liquids are not expected to be stable to all perturbations, but, at most, to a finite number of them. A possible 
instability is towards the formation of a regular pattern of singlets, i.e., the creation of a valence-bond solid. Here, while 
spin-spin correlations decay to zero, thus not showing magnetic order, singlet-singlet correlations display the signature of 
the breaking of the translational symmetry.~\cite{Balents2010}

\begin{figure}
\begin{center}
\includegraphics[width=0.35\columnwidth]{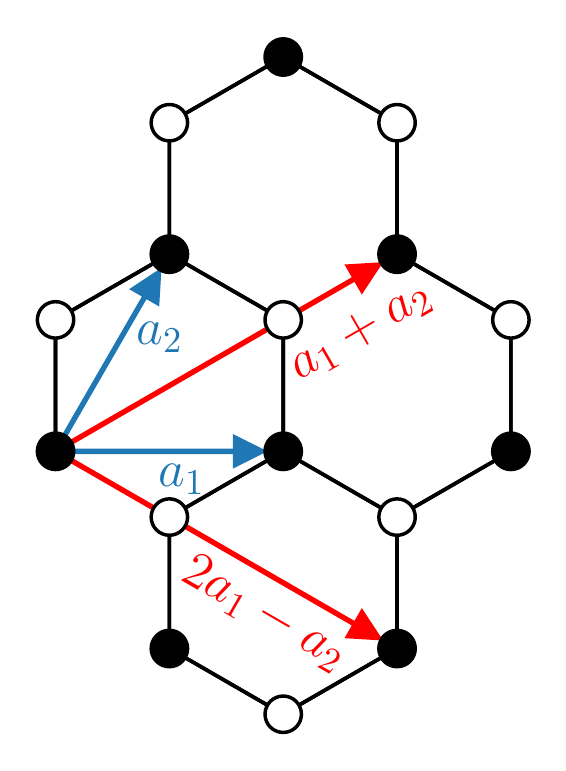}
\includegraphics[width=0.40\columnwidth]{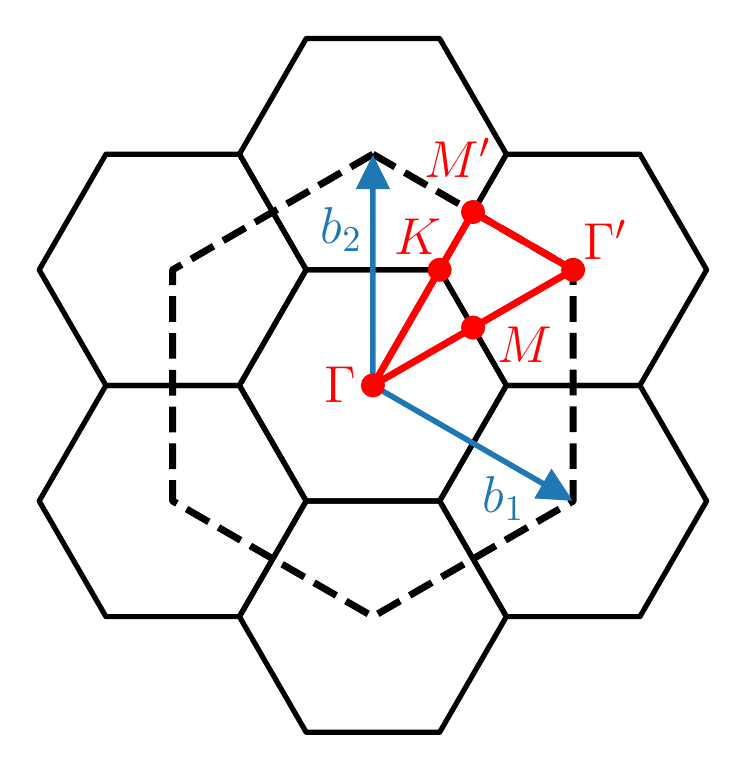}
\caption{\label{fig:latbz}
Left panel: the honeycomb lattice. The unit vectors $a_1$ and $a_2$ are represented by the blue arrows, and the sites of the 
two sublattices are represented by filled ($\delta_0$) and empty ($\delta_1$) dots. The red arrows indicate the lattice vectors 
corresponding to the periodicity of the plaquette phase. Right panel: reciprocal space of the honeycomb lattice. The blue 
arrows correspond to the reciprocal lattice vectors ${b_1=(1/\sqrt{3},-1/3)}$ and ${b_2=(0,2/3)}$. The solid black lines delimit 
the Brillouin zone (which is periodically repeated in the picture), while the dashed black lines depicts the extended Brillouin 
zone. The red line represents the path in reciprocal space along which the dynamical structure factor is displayed (except for 
Fig.~\ref{fig:dimer040}).}
\end{center}
\end{figure}

In the recent years, there have been important achievements in both the theoretical characterization of quantum spin liquids and in 
their identification in materials and spin models.~\cite{Mila2011} In this work, we focus our attention on the spin-1/2 $J_1-J_2$ 
Heisenberg model on the honeycomb lattice, which has been the subject of several investigations in the recent past. The Hamiltonian 
of the model contains antiferromagnetic exchange interactions between first- and second-neighboring sites, whose coupling strengths 
are denoted by $J_1$ and $J_2$, respectively:
\begin{equation}\label{eq:j1j2_ham}
\mathcal{H}=J_1 \sum_{\langle i,j\rangle} \mathbf{S}_i \cdot \mathbf{S}_j +
J_2 \sum_{\langle\langle i,j \rangle\rangle} \mathbf{S}_i \cdot \mathbf{S}_j.
\end{equation}
The honeycomb lattice is formed by a triangular Bravais lattice with unit vectors $a_1=(\sqrt{3},0)$ and $a_2=(\sqrt{3}/2,3/2)$, 
and a unit cell containing two sites sitting at the positions $\delta_0=(0,0)$ and $\delta_1=(0,1)$, see Fig.~\ref{fig:latbz}. 
We label the lattice positions as $i=(R_i,\alpha_i)$, where $R_i$ is the Bravais vector corresponding to the unit cell of the site 
$i$ ($R_i=n_i a_1+m_i a_2$, $n_i$ and $m_i$ being integers) and $\alpha_i$ indicates the sublattice shift ($\delta_{\alpha_i}$, 
$\alpha_i=0,1$). The first-neighbor exchange $J_1$ couples spins belonging to different sublattices, while the second-neighbor 
exchange $J_2$ involves spins sitting on the same sublattice.

%%%%%%%%%%%%%%%%%%%%%%%%%%%%%%%%%%%%%%%%%%%%%%%%%%%%%%%%%%%%%%%%%%%%%%%%%%%%%%%%%%%%%%%%%%%%%%%%%%%%%%%%%%%%%%

\begin{figure}
\begin{center}
\includegraphics[width=\columnwidth]{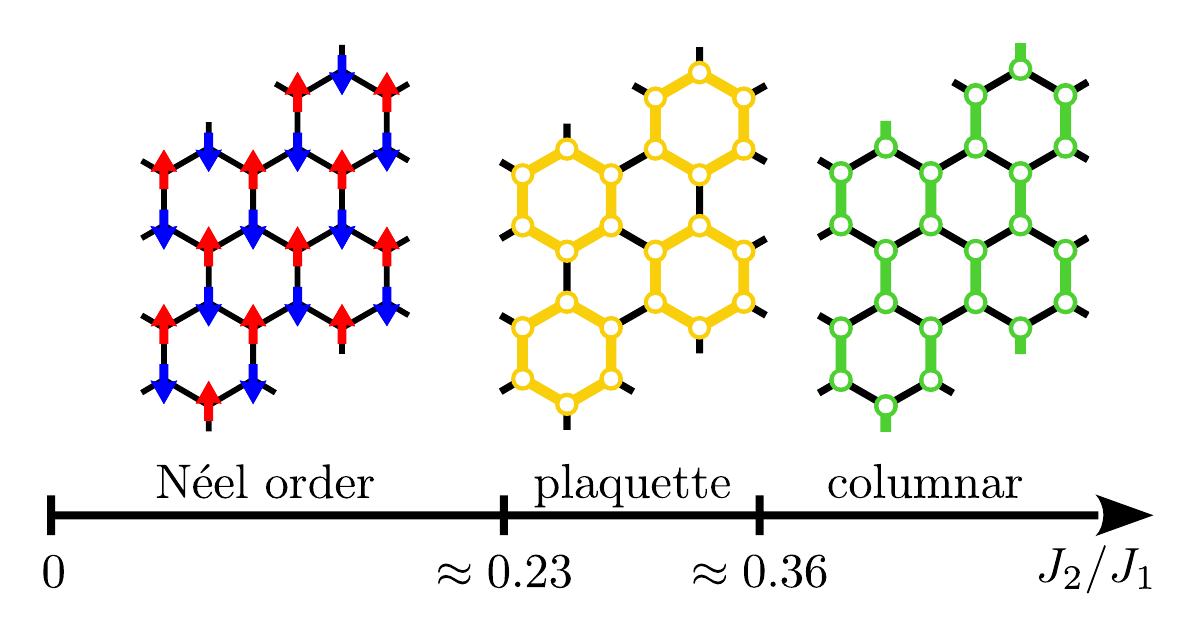}
\caption{\label{fig:phasediag}
Schematic illustration of the variational phase diagram of the $J_1-J_2$ antiferromagnetic Heisenberg model on the honeycomb 
lattice.~\cite{Ferrari2017}}
\end{center}
\end{figure}

\begin{figure}
\begin{center}
\includegraphics[width=\columnwidth]{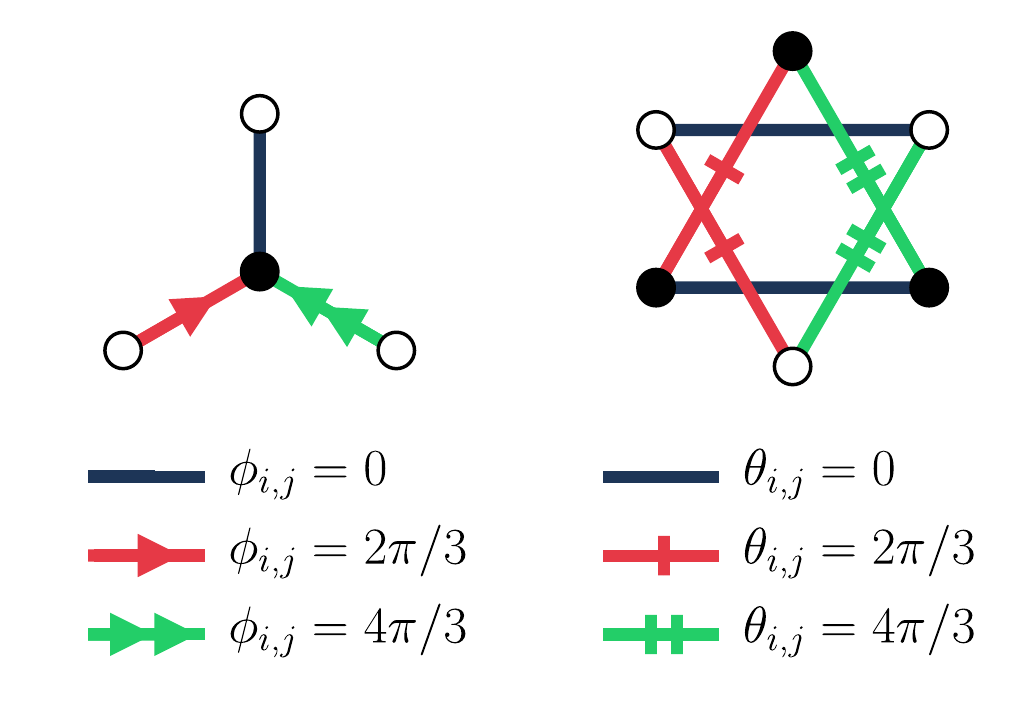}
\caption{\label{fig:phases}
Schematic illustration of the $d \pm id$ spin liquid state, where $\phi_{ij}$ and $\theta_{ij}$ are the complex phases of first-neighbor 
hopping and second-neighbor pairing, respectively. The direction of the arrows ($i\rightarrow j$) indicates the convention of phases for 
the hopping terms.}
\end{center}
\end{figure}
%%%%%%%%%%%%%%%%%%%%%%%%%%%%%%%%%%%%%%%%%%%%%%%%%%%%%%%%%%%%%%%%%%%%%%%%%%%%%%%%%%%%%%%%%%%%%%%%%%%%%%%%%%%%%%

Among the previous works, we mention semiclassical~\cite{Fouet2001,Mulder2010} and Schwinger boson approaches,~\cite{Merino2018}
techniques based on variational wave functions,~\cite{Clark2011,Mezzacapo2012,diCiolo2014} coupled-cluster calculations,~\cite{Bishop2013}
exact diagonalizations,~\cite{Albuquerque2011} and various renormalization group methods.~\cite{Reuther2011,Zhu2013,Gong2013,Ganesh2013} 
In a recent investigation,~\cite{Ferrari2017} we performed a systematic analysis of Gutzwiller-projected fermionic wave functions for 
$J_2/J_1 < 0.5$, by means of the so-called projective symmetry group (PSG) classification that has been performed by Lu and Ran.~\cite{Lu2011} 
The resulting ground-state phase diagram is reported in Fig.~\ref{fig:phasediag}. It shows three phases: (i) a magnetically ordered 
antiferromagnet with pitch vector $Q=(0,0)$ (and opposite spins on the two sites of the unit cell) for $J_2/J_1 \lesssim 0.23$; 
(ii) a magnetically disordered phase (three-fold degenerate) with plaquette order for $0.23 \lesssim J_2/J_1 \lesssim 0.36$; and (iii) 
another magnetically disordered phase (again three-fold degenerate) with dimer order for $J_2/J_1 \gtrsim 0.36$ (up to the value 
$J_2/J_1=0.5$, where we ended our investigations). In addition, a spin-liquid state was found to have competing energies in the 
proximity of the phase transition leading to the plaquette phase.~\cite{Ferrari2017}

In order to go beyond what has been done in Ref.~\cite{Ferrari2017} and assess the low-energy excitations of the system, we compute 
the dynamical structure factor within the variational scheme that has been proposed by Li and Yang.~\cite{Li2010} By using this approach, 
we recently studied the dynamical spectra of a few frustrated Heisenberg models, both in one~\cite{Ferrari2018a} and two 
dimensions~\cite{Ferrari2018b,Ferrari2019}. In the latter case (with square and triangular lattices) the variational approximation 
of the ground state neither yields dimer nor plaquette order, but instead describes a (continuous) transition between magnetically
ordered and spin-liquid phases. In this respect, the $J_1-J_2$ model on the honeycomb lattice is interesting because it provides an 
example in which the actual ground state of the system has plaquette/dimer order; in addition, the existence of a spin-liquid state 
with competing energies would allow us to directly compare the dynamical responses of these different phases.

\section{The variational approach}

In this work, we employ a variational Monte Carlo method based on Gutzwiller-projected fermionic wave functions. The variational 
{\it Ans\"atze} $|\Psi_0\rangle$ are defined by introducing an auxiliary Hamiltonian of Abrikosov fermions (${\cal H}_{0}$), and 
projecting its ground state $|\Phi_0\rangle$ onto the Hilbert space of spins:
\begin{equation}
|\Psi_0\rangle = \mathcal{P}_{S^z_{tot}} \mathcal{P}_G |\Phi_0 \rangle
\end{equation}
Here, in addition to the Gutzwiller projector, ${\mathcal{P}_G=\prod_i (n_{i,\uparrow}-n_{i,\downarrow})^2}$, which forces all sites 
to be singly occupied, a second projector is applied, $\mathcal{P}_{S^z_{tot}}$, which constrains the wave function to the sector of 
the Hilbert space in which the $z$-component of the total spin is zero. 

Adopting this variational scheme, the phase diagram of the $J_1-J_2$ model has been obtained.~\cite{Ferrari2017} 
For $J_2/J_1 \lesssim 0.23$, the system displays N\'eel magnetic order, and the optimal variational wave function is obtained by 
projecting the ground state of the auxiliary Hamiltonian
\begin{eqnarray}
 \mathcal{H}_0=  -t \sum_{\langle i,j \rangle}& 
 ( c_{i,\uparrow}^\dagger  c_{j,\uparrow}^\dagga + c_{i,\downarrow}^\dagger  c_{j,\downarrow}^\dagga ) + {\rm h.c.} \nonumber \\
 &+ h \sum_{i}   (-)^{\alpha_i}  ( c_{i,\uparrow}^\dagger c_{i,\downarrow}^\dagga
            +  c_{i,\downarrow}^\dagger c_{i,\uparrow}^\dagga  ),
\label{eq:ham0mag}
\end{eqnarray}
which contains a first-neighbor uniform hopping term $t$ and a fictitious N\'eel magnetic field $h$ along the $x$ direction. 
Additionally, a spin-spin Jastrow factor is included, in order to introduce transverse quantum fluctuations on top of the in-plane 
magnetically ordered {\it Ansatz}
\begin{equation}\label{eq:jastrowfactor}
\mathcal{J}_s=\exp \left ( \frac{1}{2} \sum_{i,j} v_{i,j} S^z_i S^z_j \right ).
\end{equation}
Both the parameters of the auxiliary Hamiltonian $\mathcal{H}_0$ and the pseudopotential $v_{i,j}$ defining the Jastrow factor are 
fully optimized by the stochastic reconfiguration technique.~\cite{Becca2017}

At $J_2/J_1 \approx 0.23$, the N\'eel magnetic order vanishes (together with the variational parameter $h$) and the system undergoes 
a phase transition to a magnetically disordered phase. At the transition point, a very accurate variational state is given by the 
Gutzwiller-projected free-fermion state (i.e., the ground state of the auxiliary Hamiltonian~(\ref{eq:ham0mag}) with $h=0$). For 
larger values of $J_2/J_1$ an energy gain is obtained by allowing a translational symmetry breaking in the hoppings, leading to a 
plaquette valence-bond solid phase; here, hexagonal plaquettes with a ``strong'' hopping (with amplitude $t$) are separated 
by each others by a ``weak'' hopping (with amplitude $t^\prime$), see Fig.~\ref{fig:phasediag}. Close to the transition point 
$J_2/J_1 \approx 0.23$, a $Z_2$ spin liquid {\it Ansatz} has a competitive energy. This state is described by an auxiliary Hamiltonian 
containing a first-neighbor hopping $t_{i,j}$ and a second-neighbor singlet pairing $\Delta_{i,j}$:
\begin{eqnarray}
 \mathcal{H}_0=  \sum_{\langle i,j \rangle} t_{i,j} 
 ( c_{i,\uparrow}^\dagger  & c_{j,\uparrow}^\dagga + c_{i,\downarrow}^\dagger  c_{j,\downarrow}^\dagga ) \nonumber \\
 &+  \sum_{\langle \langle i,j \rangle \rangle} \Delta_{i,j} c_{i,\downarrow} c_{j,\uparrow} +h.c.,
\end{eqnarray}
The hopping (${t_{i,j}=te^{i\phi_{i,j}}}$, ${t\in\mathbb{R}}$) and pairing (${\Delta_{i,j}=\Delta e^{i\theta_{i,j}}}$, ${\Delta\in\mathbb{R}}$) 
terms defining the above Hamiltonian have specific complex phases which are illustrated in Fig.~\ref{fig:phases}. Due to the way the pairing 
phases transform under rotation, this variational {\it Ansatz} has been dubbed $d\pm id$ spin liquid.~\cite{Ferrari2017} The wave 
function obtained by projecting the ground state of the above Hamiltonian fulfills all the symmetries of the lattice and is time-reversal 
invariant. Remarkably, also this spin-liquid wave function is unstable towards the realization of a plaquette valence-bond solid, which breaks 
the translational symmetry of the lattice. Indeed, a substantial energy gain is achieved by letting the {\it amplitudes} of the hopping terms 
assume different values for the first-neighbor bonds inside ($t$) and outside ($t^\prime$) the hexagonal plaquettes depicted in 
Fig.~\ref{fig:phasediag}. Even if the optimal values for $t$ and $t^\prime$ are sizeably different only for ${J_2/J_1 \gtrsim 0.26}$, 
a finite size scaling analysis of the plaquette order parameters suggested that the transition point should located at a lower value of 
the frustrating ratio, i.e. $J_2/J_1 \approx 0.23$.~\cite{Ferrari2017} This result is compatible with the existence of a continuous transition 
between the antiferromagnetic and the plaquette valence-bond solid phases.

Finally, at $J_2/J_1 \approx 0.36$, a first-order phase transition is observed, from the plaquette phase to a dimer valence-bond solid phase 
with columnar order. The optimal wave function for the dimer phase is obtained by starting from the auxiliary Hamiltonian of the $d\pm id$ 
spin liquid state and letting the amplitudes of the hoppings break the rotational symmetry of the lattice, see Fig.~\ref{fig:phasediag}. 
Upon optimization, a strong intra-dimer hopping $t$ and a weak inter-dimer hopping $t^\prime$ are obtained. Within our choice for the 
orientation of the columnar phase, the intra-dimer hopping $t$ is the term connecting the two sites that belong to the same unit cell.

The purpose of the present work is computing the dynamical structure factor of the $J_1-J_2$ Heisenberg model~(\ref{eq:j1j2_ham}) within 
all the different phases discussed so far. In particular, we are interested in the $z$-component of the dynamical structure factor,
whose exact expression is given by:
\begin{eqnarray}\label{eq:sqwab}
S_{\alpha,\beta}^z(q,\omega)=\sum_n \langle \Upsilon_0| S^z_{-q,\alpha} | & \Upsilon_n \rangle 
\langle \Upsilon_n| S^z_{q,\beta} | \Upsilon_0  \rangle \times \nonumber \\ 
& \times \delta(\omega-E_n+E_0).
\end{eqnarray}
Here, $|\Upsilon_0\rangle$ is the exact ground state (with energy $E_0$) and the sum runs over all the excited states $|\Upsilon_n\rangle$, 
whose corresponding energies are denoted by $E_n$. Moreover, ${S_{q,\alpha}^z=\frac{1}{\sqrt{N}}\sum_{R} e^{i q \cdot R} S_{R,\alpha}^z}$
are the Fourier-transformed spin operators for a lattice containing $N$ unit cells (i.e., $2N$ spins). Since the terms $S_{q,\alpha}^z$ 
explicitly depend on the sublattice label $\alpha$, the dynamical structure factor of Eq.~(\ref{eq:sqwab}) is a $2\times2$ matrix with 
respect to the indices $\alpha,\beta$. Then, the $z$-component of the total dynamical structure factor is obtained by taking the following 
linear combination:
\begin{equation}\label{eq:totsqw}
S^z(q,\omega) = \sum_{\alpha,\beta=0,1}  e^{iq(\delta_\alpha-\delta_\beta)} S_{\alpha,\beta}^z(q,\omega).
\end{equation}
We emphasize that the variational wave function for the N\'eel phase, defined by Eq.~\ref{eq:ham0mag}, breaks the spin rotational 
symmetry of the model and displays finite magnetic order in the $x-y$ plane. Therefore, within this phase, $S^z(q,\omega)$ 
corresponds to the {\it transverse} component of the dynamical structure factor, which shows the magnon branch.

The evaluation of the dynamical structure factor of Eq.~(\ref{eq:sqwab}) is performed by introducing a basis set of {\it projected} 
particle-hole spinon excitations. For each momentum $q$, we can define $O(N)$ triplet states:
\begin{eqnarray}\label{eq:qRab}
 |q &;R,\alpha;\beta\rangle =  \mathcal{P}_{S^z_{tot}}  \mathcal{P}_G
  \frac{1}{\sqrt{N}} \sum_{R^\prime} e^{iq \cdot R^\prime} \times \nonumber \\
  &\times\frac{1}{2} \left(c^\dagger_{R+R^\prime,\alpha,\uparrow}c^\dagga_{R^\prime,\beta,\uparrow}-
 c^\dagger_{R+R^\prime,\alpha,\downarrow}c^\dagga_{R^\prime,\beta,\downarrow}\right)|\Phi_0\rangle,
\end{eqnarray}
which are labelled by a Bravais lattice vector, $R$, and two sublattice indices, $\alpha$ and $\beta$. In the case of the 
N\'eel ordered phase, we include also the spin-spin Jastrow factor $\mathcal{J}_s$ in the definition of the above excitations.

%%%%%%%%%%%%%%%%%%%%%%%%%%%%%%%%%%%%%%%%%%%%%%%%%%%%%%%%%%%%%%%%%%%%%%%%%%%%%%%%%%%%%%%%%%%%%%%%%%%%%%%%%%%%%%
\begin{figure}
\begin{center}
\includegraphics[width=\columnwidth]{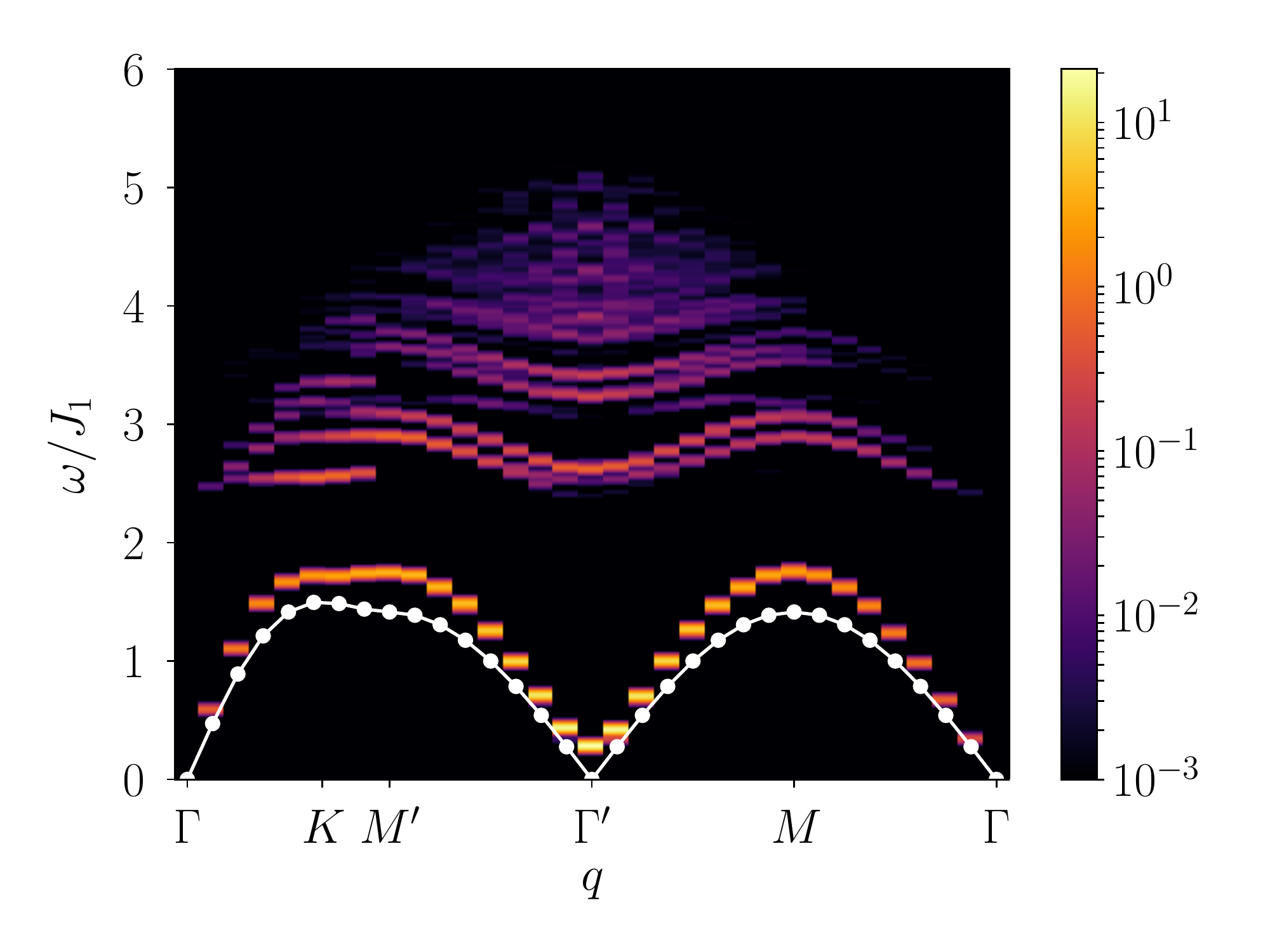}
\includegraphics[width=\columnwidth]{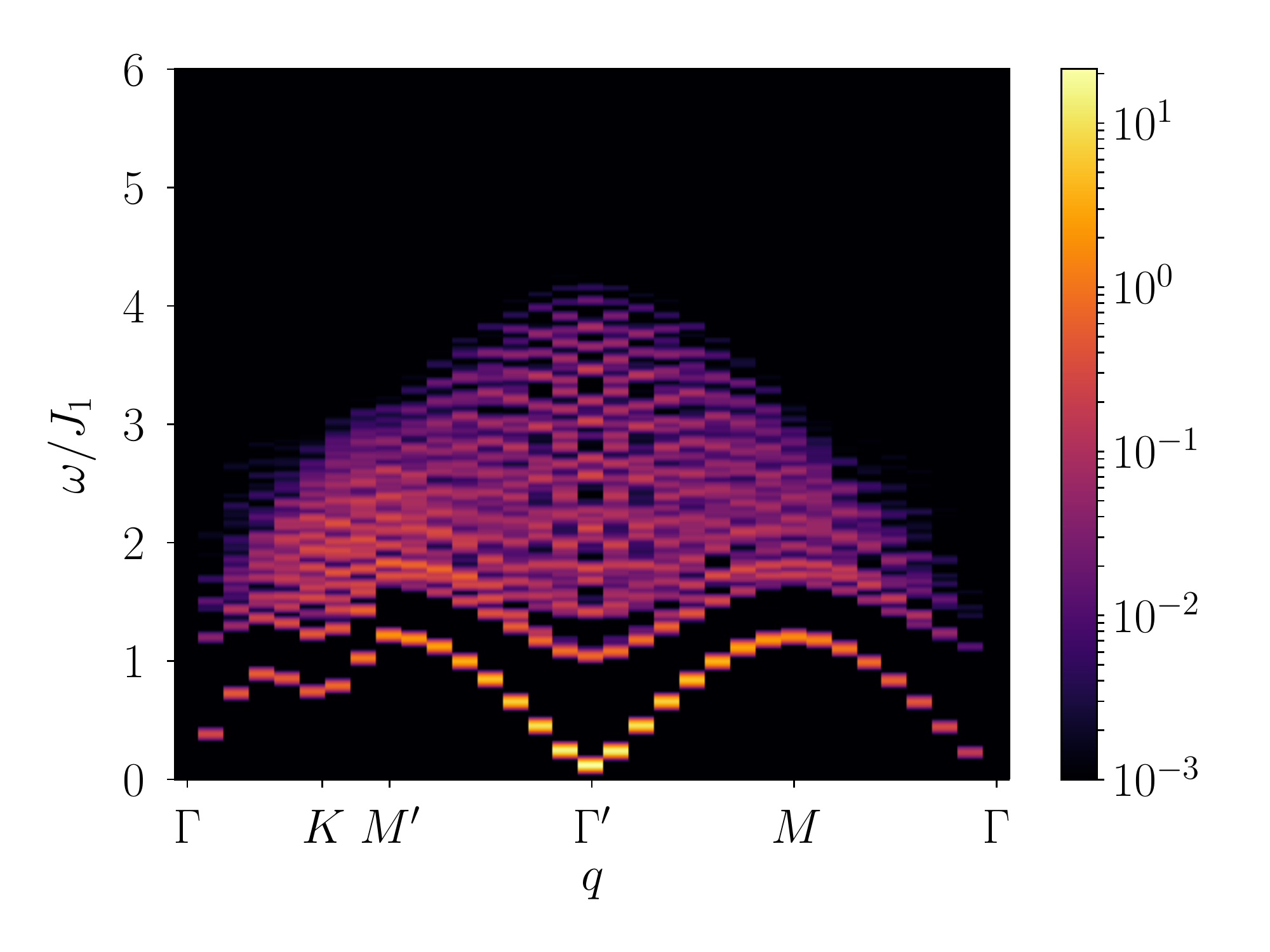}
\caption{\label{fig:neelsqw}
Dynamical structure factor of Eq.~(\ref{eq:totsqw}) for the N\'eel phase at $J_2=0$ (upper panel) and $J_2/J_1=0.15$ (lower panel). 
The calculations are performed in a clusted with $512$ sites, which is defined by the translation vectors $T_1=16 a_1$ and $T_2=16 a_2$.
The spectral functions have been convoluted with normalized Gaussians with $\sigma=0.02J_1$. The white line with dots correspond to 
the magnon dispersion of the linear spin wave theory.~\cite{Mulder2010}}
\end{center}
\end{figure}

\begin{figure}
\begin{center}
\includegraphics[width=\columnwidth]{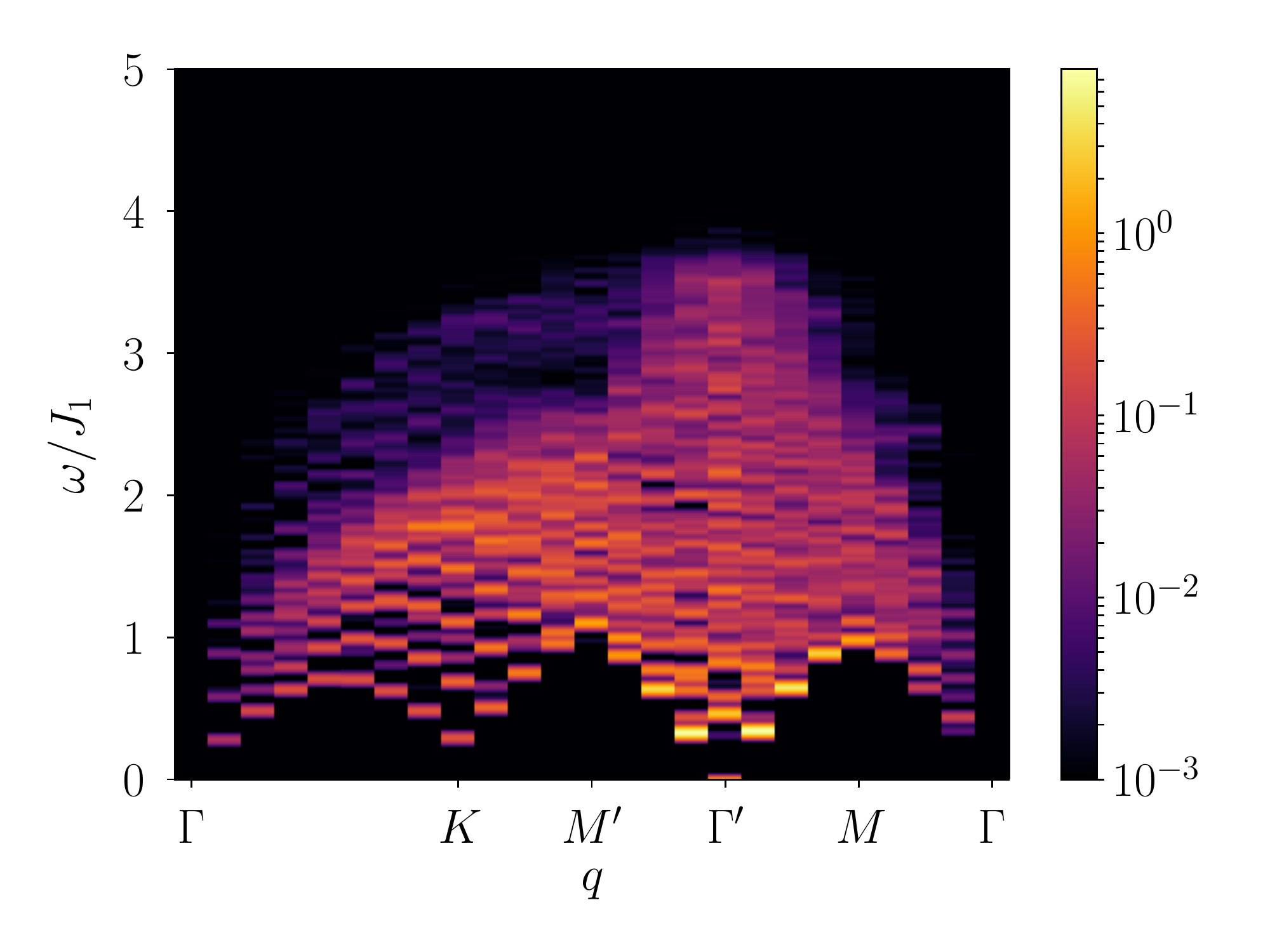}
\caption{\label{fig:sl023}
Dynamical structure factor of Eq.~(\ref{eq:totsqw}) for the $d\pm id$ spin liquid phase at $J_2/J_1=0.23$. The cluster employed for 
the calculation has $384$ sites and is defined by the translation vectors $T_1 = 8a_1 + 8a_2$ and $T_2 = -8a_1 + 16a_2$.
The spectral function has been convoluted with a normalized Gaussian with $\sigma=0.02J_1$.}
\end{center}
\end{figure}
%%%%%%%%%%%%%%%%%%%%%%%%%%%%%%%%%%%%%%%%%%%%%%%%%%%%%%%%%%%%%%%%%%%%%%%%%%%%%%%%%%%%%%%%%%%%%%%%%%%%%%%%%%%%%%

Our variational description of the excited states of the system with momentum $q$ relies on linear combinations of the elements of 
$\{|q;R,\alpha;\beta\rangle\}$:
\begin{equation}\label{eq:var_excitations}
|\Psi_n^q\rangle = \sum_{R} \sum_{\alpha,\beta} A^{n,q}_{R,\alpha;\beta} |q;R,\alpha;\beta\rangle.
\end{equation}
The optimal coefficients of the expansion are obtained by the Rayleigh-Ritz variational method, which requires the solution of the 
generalized eigenvalue problem:
\begin{eqnarray}\label{eq:general_eig_prob}
\sum_{R^\prime,\alpha^\prime,\beta^\prime} & H^q_{R,\alpha;\beta|R^\prime,\alpha^\prime;\beta^\prime} 
A^{n,q}_{R^\prime,\alpha^\prime;\beta^\prime} \nonumber \\
&= E_n^q \sum_{R^\prime,\alpha^\prime,\beta^\prime} O^q_{R,\alpha;\beta|R^\prime,\alpha^\prime;\beta^\prime} 
A^{n,q}_{R^\prime,\alpha^\prime;\beta^\prime}.
\end{eqnarray}
for each different value of $q$. Here, $H^q_{R,\alpha;\beta|R^\prime,\alpha^\prime;\beta^\prime}$ and 
$O^q_{R,\alpha;\beta|R^\prime,\alpha^\prime;\beta^\prime}$ are the Hamiltonian and overlap matrices in the basis set of excitations, namely
\begin{eqnarray}
  H^q_{R,\alpha;\beta|R^\prime,\alpha^\prime;\beta^\prime}&=\langle q;R,\alpha;\beta|{\cal H}|q;R^\prime,\alpha^\prime;\beta^\prime \rangle, \\
  O^q_{R,\alpha;\beta|R^\prime,\alpha^\prime;\beta^\prime}&=\langle q;R,\alpha;\beta|q;R^\prime,\alpha^\prime;\beta^\prime \rangle.
\end{eqnarray}
The entries of these matrices are sampled by the Monte Carlo scheme described in Ref.~\cite{Ferrari2018a}.

Once the solution of the generalized eigenvalue problem is found, the dynamical structure factor of Eq.~(\ref{eq:sqwab}) is approximated 
by taking:
\begin{eqnarray}\label{eq:dsf_tensor_approx}
S_{\alpha,\beta}^{z}(q,\omega) = \sum_n \langle \Psi_0 | S^z_{-q,\alpha} | & \Psi_{n}^q \rangle
 \langle \Psi_{n}^q | S^z_{q,\beta} | \Psi_0 \rangle \times \nonumber \\
 & \times \delta(\omega-E_{n}^q+E_0^{\rm var}),
\end{eqnarray}
where $E_0^{\rm var}$ is the ground state variational energy. Finally, the total dynamical structure factor is computed by performing
the linear combination of Eq.~(\ref{eq:totsqw}).

\section{Results}

Let us start our investigations with the unfrustrated Heisenberg model (i.e., $J_2=0$). Here, the ground state has a finite magnetization,
in which spins on the same sublattice are ferromagnetically ordered [thus leading to a periodicity with $Q=(0,0)$], while spins on
opposite sublattices are antiferromagnetically oriented. The dynamical structure factor has a periodicity on the extended Brillouin zone,
see Fig.~\ref{fig:latbz}. The results for a clusted defined by the translation vectors $T_1=16 a_1$ and $T_2=16 a_2$ (i.e., containing 
$512$ sites) are shown in Fig.~\ref{fig:neelsqw}. Here, the lowest-energy excitation for each momentum $q$ agrees quite well with the 
magnon dispersion obtained within the spin-wave approximation.~\cite{Mulder2010} In addition, our variational Monte Carlo calculations 
show a bunch of excitations with a much weaker spectral weight at energies $\omega/J_1 \gtrsim 2.5$, which are interpreted as the continuum 
generated by the magnon-magnon interactions. The presence of a finite frustrating coupling $J_2$ gives rise to a few important modifications 
of the spectral properties. For $J_2/J_1=0.15$, the magnon branch is significantly renormalized (see Fig.~\ref{fig:neelsqw}): its velocity in 
$\Gamma$ and ($\Gamma^\prime$) is reduced with respect to the unfrustrated case; most importantly, a roton-like mode around $K$ (the corner 
of the Brilloun zone) appears. In addition, a broad continuum intensifies its intensity, at the expenses of the magnon branch.

For $J_2/J_1 \approx 0.23$, the fictitious magnetic field $h$ of the auxiliary Hamiltonian~(\ref{eq:ham0mag}) vanishes and the best
variational wave function is no longer magnetically ordered. Close to this transition point, the optimal state is given by the $d\pm id$
state described above. We would like to emphasize that, for ${0.23 \lesssim J_2/J_1 \lesssim 0.26}$, we cannot stabilize any state with 
finite valence-bond order, but we cannot exclude that such order is indeed present, although invisible to our numerical simulations 
on finite clusters. The result of the dynamical structure factor in this regime is reported in Fig.~\ref{fig:sl023}. For this calculation, 
we consider a cluster defined by ${T_1 = 8a_1 + 8a_2}$ and ${T_2 = -8a_1 + 16a_2}$, in order to have a direct comparison with the results 
for the plaquette valence-bond state that are presented below. The dynamical structure factor for the spin liquid state shows a spectrum 
that is compatible with the presence of gapless excitations at $\Gamma$ ($\Gamma^\prime$) and $K$, which can be traced back to the existence 
of Dirac nodes in the band structure of the unprojected fermionic state.~\cite{Ferrari2017} The spectrum is characterized by a broad 
continuum of states that develops right above the lowest-energy excitation for each momenta. Qualitatively, these features closely resemble 
what has been already obtained for the frustrated ${J_1-J_2}$ model on the square~\cite{Ferrari2018b} and triangular~\cite{Ferrari2019} 
lattices. The large number of excitations with similar spectral weight, which form a remarkable spectral continuum even on this relatively 
small cluster, strongly suggest the existence of (almost or completely) delocalized ${S=1/2}$ spinon excitations, at least in proximity of 
${J_2/J_1 \approx 0.23}$. 

%%%%%%%%%%%%%%%%%%%%%%%%%%%%%%%%%%%%%%%%%%%%%%%%%%%%%%%%%%%%%%%%%%%%%%%%%%%%%%%%%%%%%%%%%%%%%%%%%%%%%%%%%%%%%%
\begin{figure}
\begin{center}
\includegraphics[width=\columnwidth]{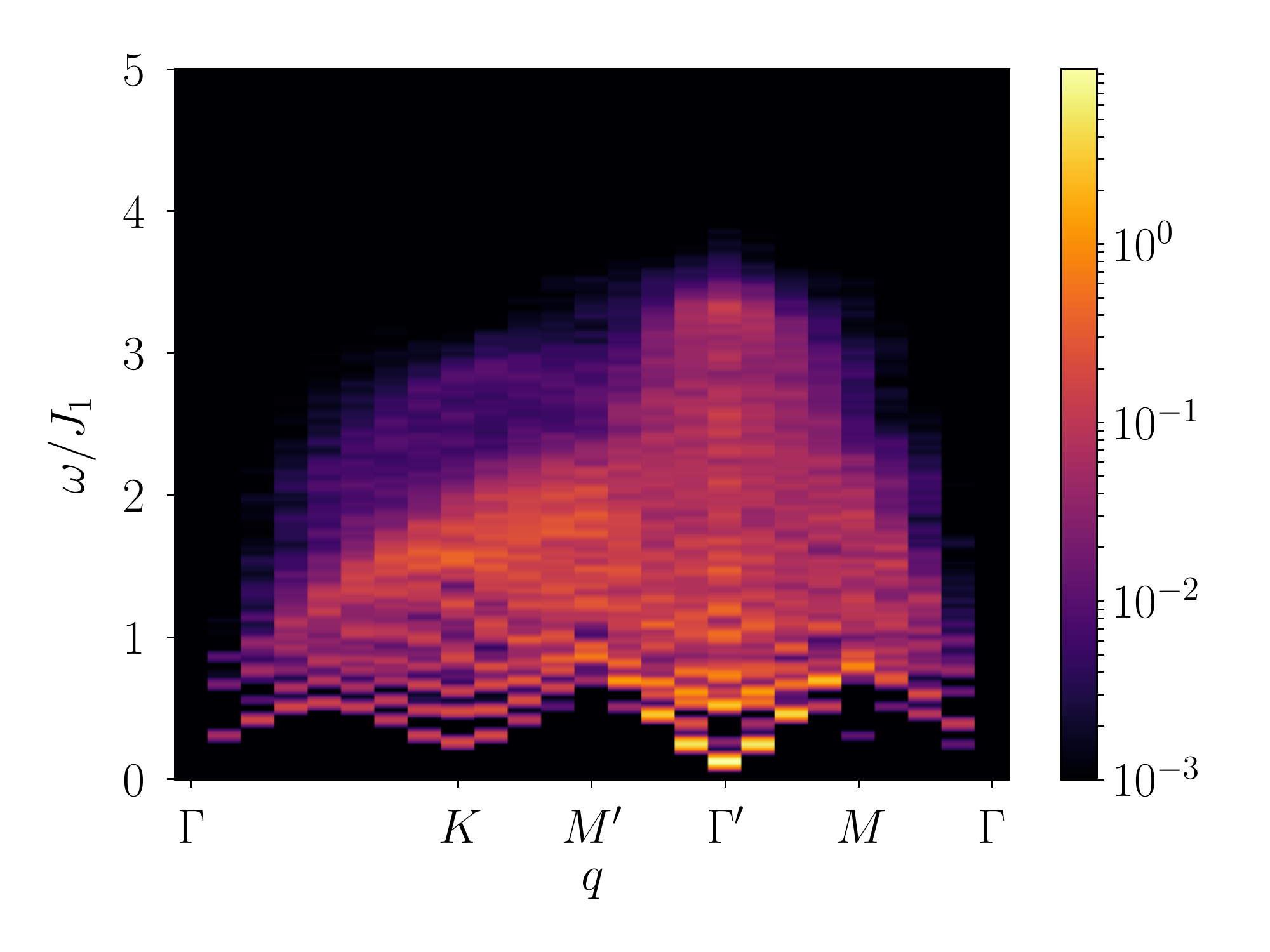}
\includegraphics[width=\columnwidth]{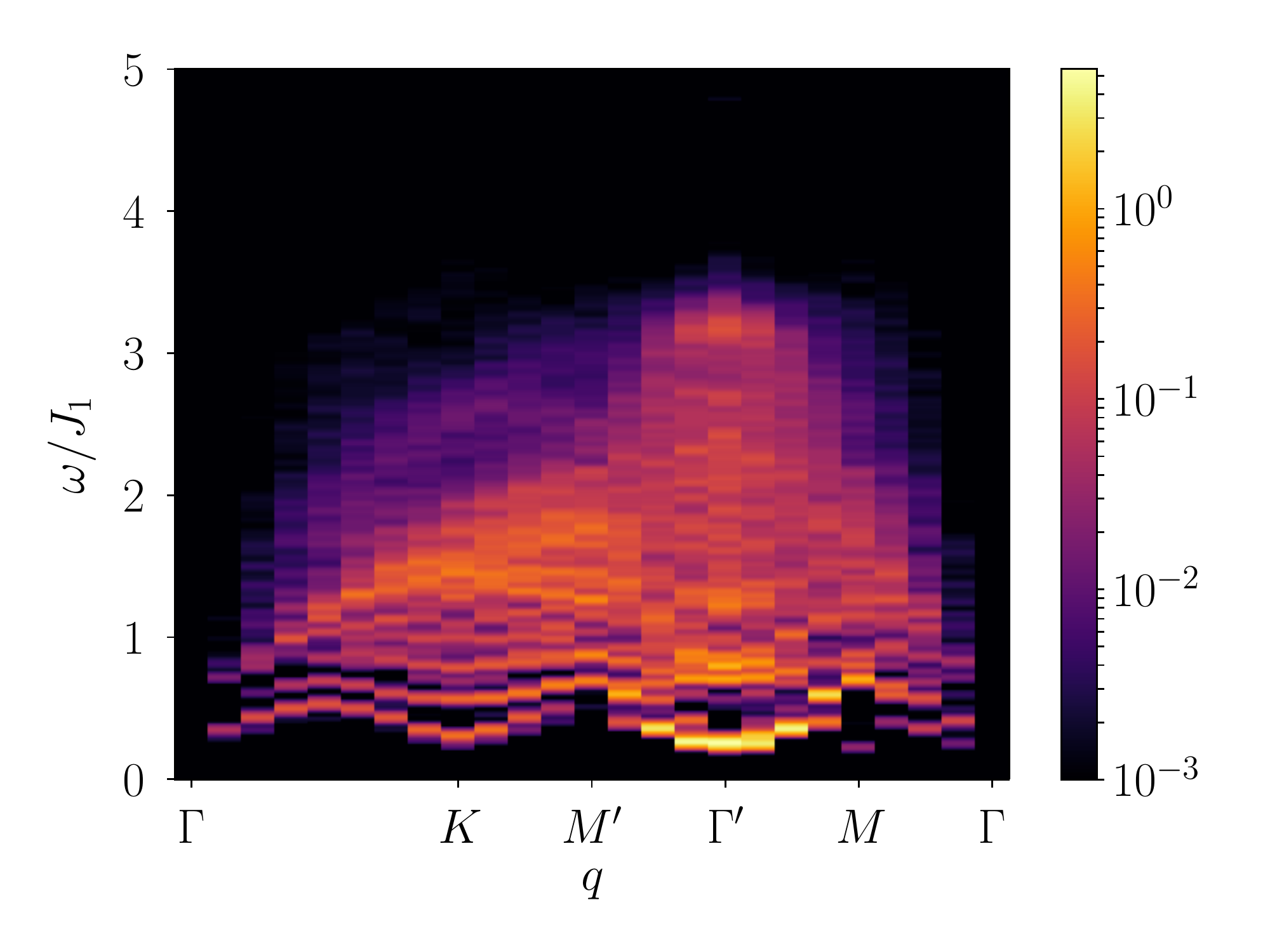}
\caption{\label{fig:plaqsqw}
Dynamical structure factor of Eq.~(\ref{eq:totsqw}) for the plaquette valence-bond solid phase at $J_2/J_1=0.30$ (upper panel) and 
$J_2/J_1=0.35$ (lower panel). The cluster employed for the calculation has $384$ sites and is defined by the translation vectors 
$T_1 = 8a_1 + 8a_2$ and $T_2 = -8a_1 + 16a_2$. The spectral functions have been convoluted with normalized Gaussians with $\sigma=0.02J_1$.}
\end{center}
\end{figure}

\begin{figure*}
\begin{center}
\includegraphics[width=1.52\columnwidth]{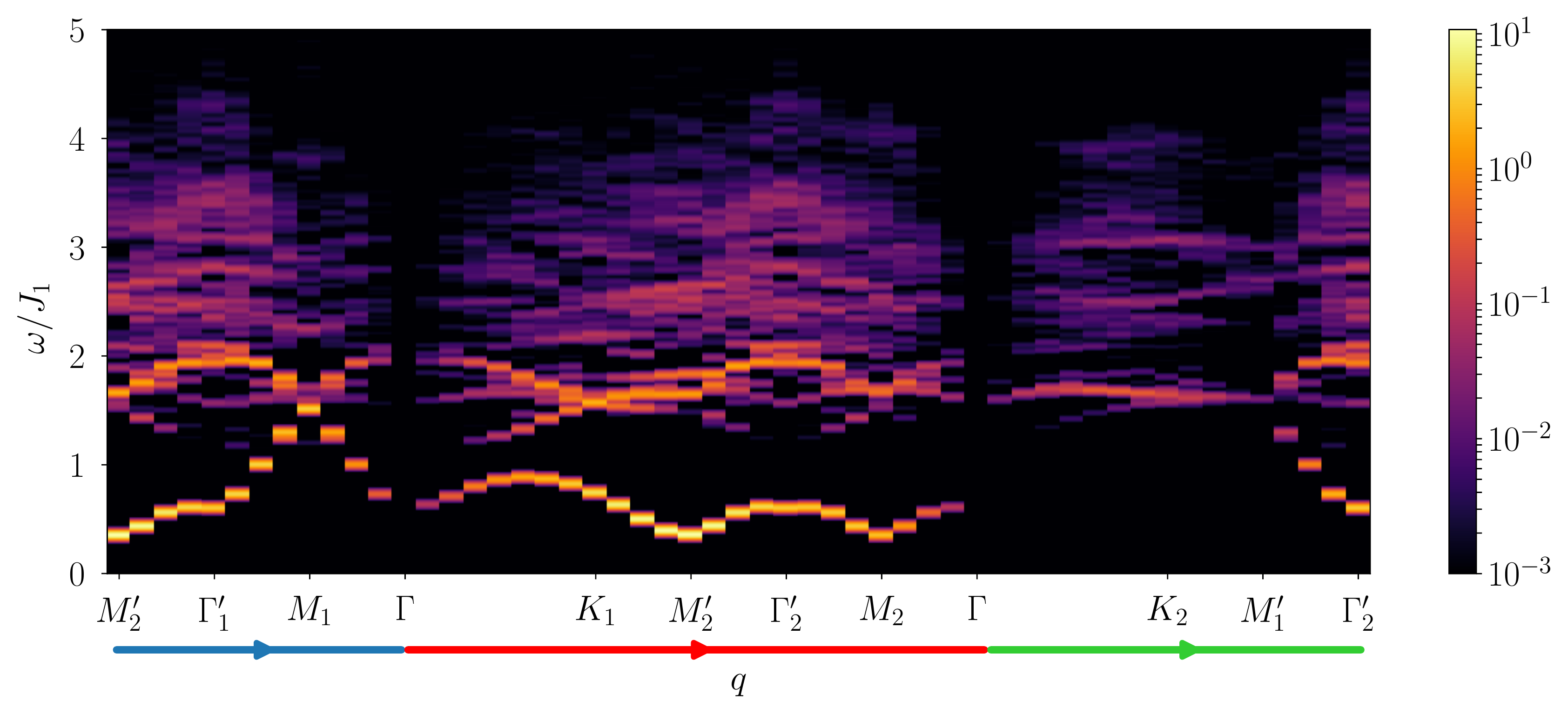}\hfill
\includegraphics[width=0.54\columnwidth]{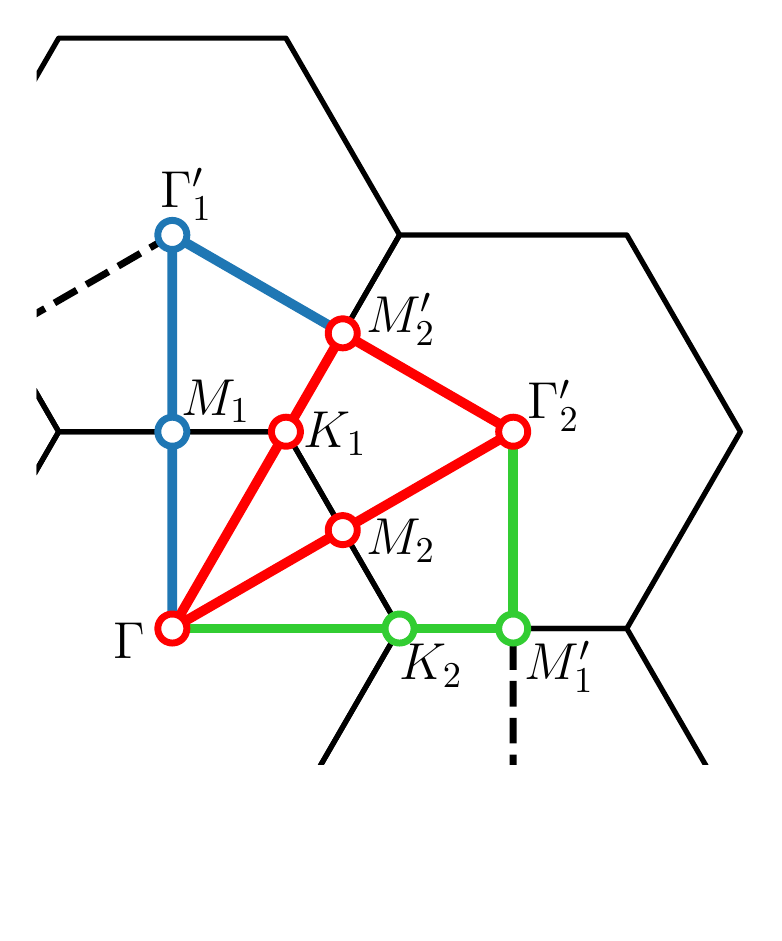}
\caption{\label{fig:dimer040}
Dynamical structure factor of Eq.~(\ref{eq:totsqw}) for the dimer valence-bond solid at $J_2/J_1=0.40$. The cluster employed for the 
calculation has $384$ sites and is defined by the translation vectors $T_1 = 8a_1 + 8a_2$ and $T_2 = -8a_1 + 16a_2$. The spectral 
function has been convoluted with a normalized Gaussian with $\sigma=0.02J_1$. The figure on the right illustrates the path in 
the Brillouin zone which has been considered.}
\end{center}
\end{figure*}
%%%%%%%%%%%%%%%%%%%%%%%%%%%%%%%%%%%%%%%%%%%%%%%%%%%%%%%%%%%%%%%%%%%%%%%%%%%%%%%%%%%%%%%%%%%%%%%%%%%%%%%%%%%%%%

At variance with the case of the square and triangular lattice models, by further increasing the ratio $J_2/J_1$, a plaquette valence-bond 
order can be stabilized (on top of the $d\pm id$ structure of hoppings and pairings). The dynamical structure factor for $J_2/J_1=0.3$ and 
$0.35$ is shown in Fig.~\ref{fig:plaqsqw} for the same cluster as the one considered for the $d\pm id$ state at $J_2/J_1=0.23$. Here, a gap 
opens at both $\Gamma$ ($\Gamma^\prime$) and $K$, and the dispersion of the lowest-energy excitation becomes ``rounded'' around these points, 
in contrast to the case of $J_2/J_1=0.23$, where Dirac-like excitations are visible. The presence of a finite gap is particularly evident 
for $J_2/J_1=0.35$ (i.e., well inside the plaquette phase): here, a sharp (triplon) mode is present around $\Gamma^\prime$, while its weight 
decreases around $K$ and goes to zero at $\Gamma$. The concomitant presence of a broad continuum may suggest that nearly-deconfined 
excitations may persist, at relatively large energies, also in the plaquette phase. 

The present results for the dynamical structure factor confirm the possibility of a {\it continuous} phase transition at $J_2/J_1\approx 0.23$, 
separating the N\'eel and the plaquette phases.~\cite{Ganesh2013} In this regard, our spectra share some similarities with the ones presented 
in Ref.~\cite{Ma2018} for an easy-plane $J-Q$ model on the square lattice. This model is characterized by a quantum phase transition between 
an antiferromagnetic phase, whose spectrum is gapless at ${q=(0,0)}$ and ${q=(\pi,\pi)}$, and a valence-bond solid phase, having a gapped spectrum.
At the transition point, the dynamical structure factor displays gapless excitations not only at $q=(0,0)$ and $q=(\pi,\pi)$, but also at 
$q=(0,\pi)$ and $q=(\pi,0)$, together with an extended continuum of states. These spectral features are interpreted as signatures of the 
presence of a deconfined quantum critical point.~\cite{Senthil2004} Analogously, in the present study, we observe a transition from the N\'eel 
state with gapless points at $\Gamma$ and $\Gamma^\prime$, to the plaquette phase, whose spectrum is fully gapped. Close to the transition 
point, $J_2/J_1\approx 0.23$, the spectral gap closes at the $K$ point and a diffuse spectral signal dominates the dynamical structure factor.

Finally, for $J_2/J_1 \approx 0.36$, a different valence-bond solid is stabilized as the best variational state. Here, a columnar dimer order 
is present, with strong bonds along one of the three nearest-neighbor distances. The transition between the states with plaquette and dimer 
orders is first order, as corroborated by the abrupt change of the spectral features. In Fig.~\ref{fig:dimer040} we report the dynamical 
structure factor of the dimer phase at  $J_2/J_1=0.4$. In this case, a different path in $q$-space is considered, in order to emphasize the 
rotational-symmetry breaking. Indeed, the lowest-energy (triplon) excitation has a vanishing weight when the $q$-vectors are transverse to
the orientation of the dimers in real space. For example, by taking the valence-bond state as in Fig.~\ref{fig:phasediag}, the variational 
state breaks rotations but not reflections with respect to the $x$ and $y$ axes, and thus the dynamical structure factor is symmetric for 
reflections with respect to $q_x=0$ and $q_y=0$. For this choice, the spectral signal vanishes along the line $\Gamma-M_1^\prime$, while 
it is quite strong in the symmetry-related path in the extended Brillouin zone, i.e. $\Gamma-M_2^\prime$. Apart from that, we would like to 
remark another important difference with respect to the plaquette phase. In fact, while in the latter case there is a visible continuum that 
starts right above the triplon mode, here the continuum has a very weak intensity at low energies, suggesting that the triplon mode is a 
well-defined bound state, which lies below the bottom of the continuum.

\section{Conclusions}

In this work, we have investigated the spectral properties of the ${J_1-J_2}$ Heisenberg model on the honeycomb lattice. A variational method 
based on Gutzwiller-projected fermionic states has been employed to track the changes of  the dynamical structure factor as a function of the
frustrating ratio ${J_2/J_1}$. In the unfrustrated regime of the model (${J_2=0}$), the variational spectra reproduce the magnon mode of the 
N\'eel phase, having gapless excitations at $\Gamma$ (and $\Gamma^\prime$), together with a continuum of two-magnon states at higher energies. 
Remaining in the magnetically ordered phase, we observe that the dispersion of the magnon branch at the $K$ point bends downwards when a 
finite second-neighbor coupling $J_2$ is included in the Hamiltonian, thus giving rise to the onset of a roton-like mode. 

At $J_2/J_1\approx 0.23$, the system undergoes a continuous transition to a plaquette valence-bond solid phase, which is characterized by a 
fully gapped spectrum. In the proximity of the transition, the dynamical structure factor exhibits a gap closing of the roton mode at the 
$K$ point, together with the appearance of a broad continuum of states. In analogy to what has been observed in Ref.~\cite{Ma2018}, these 
features may be interpreted as the hallmark of a quantum critical point in which $S=1/2$ spinon excitations are deconfined. The plaquette
phase of the model survives up to $J_2/J_1\approx 0.36$, where a first order phase transition to a dimer valence-bond solid state takes place. 
Within this phase, the spectrum of the system displays a sharp (gapped) triplon mode at low energies, whose spectral weight vanishes for
$q$-vectors that are transverse to the orientation of the dimers in real space.

\section*{References}

\bibliographystyle{iopart-num}
\bibliography{biblio}

\end{document}